\documentclass[12pt,a4paper]{article}
\usepackage[utf8]{inputenc}
\usepackage[T1]{fontenc}
\usepackage{graphicx}
\usepackage[margin=1in]{geometry}
\usepackage{float}
\usepackage{amsthm}
\usepackage{amsmath}
\usepackage{amsfonts}
\usepackage{amssymb}

\newcommand{\AES}{\mathsf{AES}}
\newcommand{\GCM}{\mathsf{GCM}}
\newcommand{\SIV}{\mathsf{SIV}}
\newcommand{\AESGCMSIV}{\AES\text{-}\GCM\text{-}\SIV}
\newcommand{\HKDF}{\mathsf{HKDF}}
\newcommand{\Argon}{\mathsf{Argon2id}}
\newcommand{\REV}{\mathsf{REV}}

\newcommand{\Encode}{\textsf{Encode}}

\usepackage{multirow}
\usepackage{algorithm}
\usepackage{algpseudocode}
\usepackage{makecell}
\usepackage{tabularx}
\usepackage{booktabs}
\usepackage{caption}
\usepackage{hyperref}
\usepackage{url}
\usepackage{tikz}

\newtheorem{claim}{Claim}

\usetikzlibrary{shapes.geometric, arrows.meta, positioning, fit, calc, backgrounds, shadows}

\tikzset{
  basicbox/.style = {draw, rounded corners, minimum height=1cm, align=center, font=\small},
  input/.style = {basicbox, fill=green!10, text width=2.2cm},
  process/.style = {basicbox, rectangle, fill=blue!10, text width=3.2cm, minimum height=1.4cm},
  secret/.style = {basicbox, fill=red!20, dashed, text width=4.5cm, font=\bfseries},
  artifact/.style = {basicbox, fill=yellow!20, double, text width=3.5cm},
  pqcbox/.style = {basicbox, dashed, fill=purple!10, text width=2cm},
  policy/.style = {basicbox, dashed, draw=orange!80, fill=orange!5, text width=3cm, font=\footnotesize\itshape},
  arrow/.style = {thick, ->, >=Stealth, rounded corners},
  lbl/.style = {fill=white, inner sep=2pt, font=\scriptsize, align=center, text=black!80}
}

\title{ACE-GF: A Generative Framework for Atomic Cryptographic Entities}
\author{
  \texttt{Jian Sheng Wang} \\
  \texttt{jason@acechain.io} \\
  Montreal, Canada
}
\date{November 24, 2026}

\begin{document}

\maketitle

%%%%%%%%%%%%%%%%%%%%%%%%%%%%%%%%%%%%%%%%%%%%%%%%%%%%%%%%%%%%
%                         Abstract                         %
%%%%%%%%%%%%%%%%%%%%%%%%%%%%%%%%%%%%%%%%%%%%%%%%%%%%%%%%%%%%

\begin{abstract}
Autonomous digital entities require deterministic identity mechanisms that avoid persistent storage of high-value master secrets, while supporting credential rotation and cryptographic agility across heterogeneous systems. Existing deterministic key hierarchies and centralized key management systems typically rely on long-lived root secrets, introducing structural single points of failure and complicating lifecycle management.

We present \textbf{ACE-GF} (Atomic Cryptographic Entity Generative Framework), a seed-storage-free identity construction that enables deterministic and context-isolated key derivation without storing any master secret at rest. The construction reconstructs an identity root ephemerally in memory from a sealed artifact and authorization credentials, using misuse-resistant authenticated encryption together with standard key derivation primitives. Derived keys are generated via HKDF with explicit context encoding, which isolates keys across curves and application domains.

The design supports stateless credential rotation, authorization-bound revocation, and non-disruptive migration toward post-quantum cryptographic domains. Its Argon2id cost parameters can be tuned to the device, so deterministic identity reconstruction remains practical on hardware ranging from high-performance servers to low-power IoT nodes, under the same security model.

This work builds upon the conceptual framework introduced in MSCIKDF~\cite{mscikdf}, which identified the core design goals for multi-curve, context-isolated, PQC-pluggable identity but did not provide a concrete construction. A formal protocol specification of ACE-GF has been submitted as an IETF Internet-Draft~\cite{draft-acegf}.
\end{abstract}

%%%%%%%%%%%%%%%%%%%%%%%%%%%%%%%%%%%%%%%%%%%%%%%%%%%%%%%%%%%%
%                   1. Introduction                        %
%%%%%%%%%%%%%%%%%%%%%%%%%%%%%%%%%%%%%%%%%%%%%%%%%%%%%%%%%%%%

\section{Introduction}

\subsection{Background and Motivation: Inherent Risks of Master Seed Storage}

Deterministic key management schemes, such as BIP-32~\cite{bip32} and BIP-39~\cite{bip39}, have greatly simplified key backup and recovery by deriving large collections of cryptographic keys from a single high-entropy master seed. Despite their practical success, these schemes rely on long-term persistent storage of the master seed. Whether it is kept as a mnemonic backup (exposed to physical compromise, duplication, and operational error) or inside a hardware security module (HSM, tied to hardware trust anchors and vendors), the master seed is a single point of failure (SPOF). Once compromised, all past and future keys derived from the hierarchy are irreversibly exposed.

This storage-centric design is particularly ill-suited for autonomous digital entities (ADEs), such as AI agents, decentralized services, and large-scale IoT deployments. These entities often operate without centralized trust anchors and require identity mechanisms that support deterministic reconstruction, credential lifecycle management, and isolation across cryptographic contexts. For such systems, the persistent storage of high-value secrets introduces operational fragility and amplifies the impact of compromise. What these systems need is an identity construction that decouples deterministic identity from long-term secret storage.

\subsection{Our Contributions}

In this work, we present \emph{ACE-GF}, a seed-storage-free identity construction designed to address the above challenges. ACE-GF introduces no new cryptographic primitives; it composes standard building blocks into an identity mechanism with the following properties:

\begin{itemize}
  \item \textbf{Seed-Storage-Free Operation.}
  The identity root, referred to as the Root Entropy Value (REV), exists only ephemerally in memory during execution and is never persistently stored or exported.

  \item \textbf{Deterministic Reconstruction.}
  The REV is deterministically reconstructed from a sealed artifact and authorization credentials, enabling reliable identity recovery without storing any master secret at rest.

  \item \textbf{Stateless Credential Rotation.}
  Authorization credentials can be updated arbitrarily without modifying or re-deriving the underlying identity root.

  \item \textbf{Stateless Authorization Revocation.}
  Future authorization-bound access can be revoked by removing credential components, without maintaining server-side state or key hierarchies; this does not undo a prior exposure of the identity root.

  \item \textbf{Strict Context Isolation and Cryptographic Agility.}
  Cryptographic keys are derived with explicit context encoding, so keys for different algorithms and application domains are independent of one another, and post-quantum cryptographic (PQC) algorithms can be added under new contexts without disrupting existing keys.
\end{itemize}

In addition, this work makes the following contributions:
(i) it introduces an abstract model for seed-storage-free identity constructions (SSF-ID) to reason about system behavior and security goals; and
(ii) it presents a concrete instantiation based only on well-established cryptographic primitives, together with supporting security arguments in a game-based model.

\begin{figure}[t]
\centering
\begin{tikzpicture}[
    font=\scriptsize,
    every node/.style={align=center}
]

% =================================================
% LEFT: Key-centric architecture
% =================================================
\begin{scope}[scale=0.7, transform shape]

% Outer frame
\draw (0,0) rectangle (9.0,8);

% Bottom layer
\draw (0,0) rectangle (9.0,3.8);
\node[font=\small\bfseries] at (4.5,3.4) {Private Keys / Seed Material};

\draw[dashed, rounded corners] (0.4,0.4) rectangle (8.6,2.6);
\node[font=\scriptsize, text width=7.8cm] at (4.5,1.5) {
PEM / DER \quad PKCS\#8 / PKCS\#12 \\
Keystore Files \quad OS Secure Storage \\
Hardware Slots (HSM / TPM / Enclave) \\
Mnemonic Phrases \quad Paper Backups
};

% Middle layer
\draw (0,3.8) rectangle (2.8,5.0);
\node at (1.4,4.4) {KMS};

\draw (2.8,3.8) rectangle (6.2,5.0);
\node at (4.5,4.4) {OpenSSL /\\ Java KeyStore /\\ etc.};

\draw (6.2,3.8) rectangle (9.0,5.0);
\node at (7.6,4.4) {Crypto Wallets};

% Top layer
\draw (0,5.0) rectangle (9.0,8);
\node[font=\small\bfseries] at (4.5,7.4) {Applications \& Systems};

\node[font=\scriptsize, text width=7.7cm] at (4.5,6.2) {
Applications / \quad Web Servers / \quad Asset Management Systems / \\
Cloud Services / \quad IoT Devices / \\
DevOps / CI Tools / \quad Client Software
};

\end{scope}

% =================================================
% Arrow
% =================================================
\node at (7.3,3.0) {\Huge$\Longrightarrow$};

% =================================================
% RIGHT: ACE-GF architecture
% =================================================
\begin{scope}[xshift=8.4cm, scale=0.7, transform shape]

% Outer frame
\draw (0,0) rectangle (9.0,8);

% Bottom layer
\draw (0,0) rectangle (9.0,3.8);
\node[font=\small\bfseries] at (4.5,3.4) {ACE-GF Identity Substrate};

\draw[dashed, rounded corners] (0.4,0.4) rectangle (8.6,2.6);
\node[font=\scriptsize, text width=7.8cm] at (4.5,1.5) {
Sealed Artifacts (no secret material at rest) \\
Deterministic, context-isolated key derivation \\
Across curves and cryptographic algorithms
};

% Middle layer
\draw (0,3.8) rectangle (2.8,5.0);
\node at (1.4,4.4) {KMS};

\draw (2.8,3.8) rectangle (6.2,5.0);
\node at (4.5,4.4) {OpenSSL /\\ Java KeyStore /\\ etc.};

\draw (6.2,3.8) rectangle (9.0,5.0);
\node at (7.6,4.4) {Crypto Wallets};

% Top layer
\draw (0,5.0) rectangle (9.0,8);
\node[font=\small\bfseries] at (4.5,7.4) {Applications \& Systems};

\node[font=\scriptsize, text width=7.7cm] at (4.5,6.2) {
Applications / \quad Web Servers / \quad Asset Management Systems / \\
Cloud Services / \quad IoT Devices / \\
DevOps / CI Tools / \quad Client Software
};

\end{scope}

\end{tikzpicture}
\caption{Evolution from key-centric to identity-centric security architecture.
Left: traditional systems rely on persistent private keys or seed material stored
at rest. Right: ACE-GF replaces persistent keys with an identity substrate that
derives cryptographic material ephemerally and contextually.}
\label{fig:architecture-comparison}
\end{figure}

\subsection{Paper Organization}

The remainder of this paper is organized as follows. Section~2 reviews related work in deterministic key derivation and key management systems. Section~3 introduces the cryptographic preliminaries and abstract definitions used to reason about seed-storage-free identity constructions. Section~4 presents the ACE-GF construction, followed by a security analysis in Section~5. Implementation considerations and deployment scenarios are discussed in Section~6, and Section~7 concludes the paper.

%%%%%%%%%%%%%%%%%%%%%%%%%%%%%%%%%%%%%%%%%%%%%%%%%%%%%%%%%%%%
%     2. Related Work and Comparative Analysis             %
%%%%%%%%%%%%%%%%%%%%%%%%%%%%%%%%%%%%%%%%%%%%%%%%%%%%%%%%%%%%

\section{Related Work and Comparative Analysis}

This section reviews existing approaches to deterministic key management and identity systems, and situates ACE-GF within the broader design space. We compare the design trade-offs of each approach with respect to persistence, trust assumptions, lifecycle management, and cryptographic isolation.

\subsection{Hierarchical Deterministic Key Management}

Hierarchical deterministic (HD) wallet constructions, most notably BIP-32~\cite{bip32} and BIP-39~\cite{bip39}, are the dominant approach to deterministic key management in cryptocurrency and related systems. These schemes derive large families of cryptographic keys from a single master seed, offering a simple and user-friendly backup and recovery model.

However, HD wallets assume that the master seed is stored long-term. The seed, whether stored directly, encoded as a mnemonic phrase, or protected by a passphrase, is the root of all derived keys and therefore a single point of failure. Once compromised, all keys across all derivation paths and cryptographic domains are permanently exposed.

From a systems perspective, HD constructions also provide limited support for fine-grained domain isolation and lifecycle management. All derived keys are ultimately rooted in the same entropy source and derivation function, and passphrase mechanisms in BIP-39 are primarily intended to generate alternative master keys rather than to support stateless credential rotation or authorization revocation. HD wallets therefore suit individual users with static backup requirements better than long-lived, autonomous, or multi-context digital entities.

\subsection{Centralized Key Management Services}

Centralized key management services (KMS), such as those provided by major cloud platforms, represent a different point in the design space. These systems avoid exposing raw key material to applications and support managed key rotation, access control, and auditing, typically backed by certified hardware security modules.

While effective in enterprise environments, centralized KMS solutions rely on trusted infrastructure, persistent server-side state, and explicit trust relationships with the service operator. Identity material is not deterministically reconstructible by the client alone, and availability, revocation, and policy enforcement are mediated by centralized control planes. This limits their use in decentralized systems and for autonomous entities that cannot rely on continuous trust in infrastructure.

\subsection{Hardware Security Modules and Secure Elements}

Hardware security modules (HSMs), trusted platform modules (TPMs), and secure elements provide strong physical protections for long-term secret storage and are widely used in high-assurance environments. By design, these devices retain master keys internally and restrict their use through controlled interfaces.

Hardware-bound secrets, however, are harder to port, deploy, and migrate. Identity continuity becomes tied to specific devices or vendors, and recovery or migration often requires specialized procedures. While HSMs can mitigate certain attack vectors, they do not eliminate reliance on persistent master secrets and are therefore orthogonal to approaches that seek to remove seed storage entirely. ACE-GF is designed as a software-level construction that can operate independently of specialized hardware, while remaining compatible with trusted execution environments (TEEs) when additional protection is desired.

\subsection{Prior Conceptual Work: MSCIKDF}

The conceptual foundations of ACE-GF were first explored in MSCIKDF~\cite{mscikdf}\footnote{MSCIKDF is Version~1 of this article (arXiv:2511.20505).}, which identified the core design goals for a single-root, multi-curve, context-isolated, PQC-pluggable cryptographic identity primitive with stateless secret rotation. MSCIKDF listed the target properties (context isolation, multi-curve independence, zero-linkability, and stateless rotation) but stayed at the level of requirements, without a concrete cryptographic construction, formal security model, or instantiation.

ACE-GF adds: (i) a concrete construction based only on standard primitives (AES-GCM-SIV, HKDF-SHA256, Argon2id); (ii) a separation of identity and authorization through the Seal/Unseal mechanism; (iii) an abstract model (SSF-ID) for reasoning about seed-storage-free identity; and (iv) security arguments under standard cryptographic assumptions.

\subsection{Comparative Analysis}

To clarify the design distinctions among existing approaches, Table~\ref{tab:comparison} summarizes key properties across representative identity and key management models. The comparison highlights differences in trust assumptions, persistence requirements, lifecycle capabilities, and cryptographic isolation, rather than absolute security guarantees.

\setlength{\extrarowheight}{1pt}
\begin{table}[ht]
  \centering
  \caption{Comparison of identity and key management approaches}
  \label{tab:comparison}
  {\scriptsize
  \begin{tabular}{lcccc}
    \toprule
    \textbf{Property} & \textbf{ACE-GF} & \textbf{BIP-32/39} & \textbf{Centralized KMS} & \textbf{HSM / TPM} \\
    \midrule
    \textbf{Persistent master secret at rest} & No & Yes & Yes & Yes \\
    \textbf{Deterministic identity reconstruction} & Yes & Yes & No & No \\
    \textbf{Stateless credential rotation} & Yes & No & No & No \\
    \textbf{Stateless authorization revocation} & Yes & No & No & No \\
    \textbf{Explicit cryptographic context isolation} & Yes & Limited & Yes & Limited \\
    \textbf{Post-quantum migration support$^\dagger$} & Yes & No & Limited & Limited \\
    \textbf{Dependence on centralized trust} & No & No & Yes & Yes \\
    \bottomrule
  \end{tabular}
  }
  \vspace{2pt}
  \par\raggedright
  {\footnotesize $^\dagger$\,Without re-enrollment or rekeying of existing identities. Some KMS vendors and HSM firmware support individual PQC key types, but lack a unified migration path that preserves identity continuity across classical and post-quantum domains.}
\end{table}

ACE-GF keeps deterministic identity reconstruction while removing persistent master secret storage and supporting stateless lifecycle operations. This combination matters most in decentralized and autonomous deployments, where long-term availability, independence from infrastructure trust, and controlled revocation are primary concerns.

%%%%%%%%%%%%%%%%%%%%%%%%%%%%%%%%%%%%%%%%%%%%%%%%%%%%%%%%%%%%
%       3. Preliminaries and Abstract Model                %
%%%%%%%%%%%%%%%%%%%%%%%%%%%%%%%%%%%%%%%%%%%%%%%%%%%%%%%%%%%%

\section{Preliminaries and Abstract Model}

This section introduces the cryptographic primitives and abstract definitions used throughout the paper. The definitions give a compact framework for reasoning about the behavior, security goals, and lifecycle properties of seed-storage-free identity constructions.

\subsection{Cryptographic Building Blocks}

ACE-GF is constructed entirely from well-established cryptographic primitives. We briefly review the relevant properties of these primitives and explain their role in the overall system design.

\paragraph{Misuse-Resistant Authenticated Encryption (AES-GCM-SIV).}
AES-GCM-SIV~\cite{aesgcmsiv} is an authenticated encryption scheme that provides nonce-misuse resistance. Unlike standard AES-GCM, which catastrophically fails under nonce reuse, AES-GCM-SIV preserves confidentiality even when nonces are reused, provided the encryption key remains secret. This property allows a fixed public nonce, so sealing high-entropy identity material requires no nonce state; seed-storage-free reconstruction depends on this.

\paragraph{Key Derivation with Explicit Context Encoding (HKDF-SHA256).}
HKDF-SHA256 is a HMAC-based key derivation function following the extract-then-expand paradigm~\cite{RFC5869}. In ACE-GF, HKDF is used with explicitly structured context strings supplied via the \texttt{info} parameter. Distinct context strings yield independent keys across curves, protocols, and application domains, so compromise in one context does not affect keys derived in another.

\paragraph{Memory-Hard Credential Derivation (Argon2id).}
Argon2id~\cite{argon2} is a memory-hard password-based key derivation function designed to resist offline brute-force and time--memory trade-off attacks. ACE-GF uses Argon2id only to derive sealing keys from human-chosen authorization credentials, thereby raising the computational cost of offline guessing attacks against sealed identity artifacts.

\subsection{Abstract Model for Seed-Storage-Free Identity}

To reason about the behavior and security properties of ACE-GF, we introduce an abstract model for seed-storage-free identity constructions. This model serves as a descriptive framework and does not claim cryptographic novelty.

A seed-storage-free identity scheme, denoted by $\mathcal{S}$, consists of the following polynomial-time algorithms:

\begin{itemize}
  \item $\textsf{Setup}(1^\lambda) \rightarrow \textsf{params}$:
  Initializes public system parameters for security parameter $\lambda$.

  \item $\textsf{Seal}(\textsf{params}, \textsf{Cred}) \rightarrow \textsf{SA}$:
  Takes an authorization credential $\textsf{Cred}$ and produces a sealed artifact $\textsf{SA}$ that encodes an identity root in protected form.

  \item $\textsf{Unseal}(\textsf{params}, \textsf{SA}, \textsf{Cred}) \rightarrow \textsf{REV} \cup \{\perp\}$:
  Given a sealed artifact and a candidate credential, reconstructs the root entropy value $\textsf{REV}$ if authentication succeeds, or outputs $\perp$ otherwise. The identity root exists only ephemerally during this operation.

  \item $\textsf{Derive}(\textsf{REV}, \textsf{Ctx}) \rightarrow \textsf{Key}$:
  Derives a cryptographic key from the root entropy value and an explicit context descriptor $\textsf{Ctx}$ encoding algorithm, domain, and application semantics.

  \item $\textsf{RotateCredential}(\textsf{SA}, \textsf{Cred}_{\text{old}}, \textsf{Cred}_{\text{new}}) \rightarrow \textsf{SA}' \cup \{\perp\}$:
  Updates the sealed artifact to bind the same underlying identity root to a new authorization credential, without modifying the identity itself.
\end{itemize}

\paragraph{Correctness.}
For any valid credential $\textsf{Cred}$ and sealed artifact $\textsf{SA}$ generated under $\textsf{Cred}$, the following holds:
\[
\textsf{Unseal}(\textsf{params}, \textsf{SA}, \textsf{Cred}) = \textsf{REV}.
\]
Conversely, for any incorrect credential, the unsealing operation fails with overwhelming probability.

The model separates identity (the REV) from authorization (the credential). Concrete instantiations, such as ACE-GF, specify how the sealed artifact is constructed, how credentials are bound, and how context-isolated derivation is realized using standard cryptographic primitives.

%%%%%%%%%%%%%%%%%%%%%%%%%%%%%%%%%%%%%%%%%%%%%%%%%%%%%%%%%%%%
%            4. The ACE-GF Construction                    %
%%%%%%%%%%%%%%%%%%%%%%%%%%%%%%%%%%%%%%%%%%%%%%%%%%%%%%%%%%%%

\section{The ACE-GF Construction}

This section presents the concrete construction of ACE-GF as an instantiation of the abstract seed-storage-free identity model introduced in Section~3.

\subsection{Overall Architecture}

ACE-GF is built around a high-entropy identity root, referred to as the Root Entropy Value (REV). The REV uniquely defines the cryptographic identity of an entity and serves as the sole source of entropy for all derived keys. Authorization credentials do not influence key derivation; they only protect access to the REV and never become part of the identity.

The construction therefore consists of two logically independent pipelines:
(i) an \emph{identity pipeline}, which deterministically derives cryptographic keys from the REV using context-isolated derivation; and
(ii) an \emph{authorization pipeline}, which controls access to the REV through a sealed artifact bound to credentials.
Because the pipelines are separate, credentials can be updated or revoked statelessly while the identity stays the same.

\begin{figure}[H]
  \centering
  \resizebox{\textwidth}{!}{%
  \begin{tikzpicture}[
    node distance=1.5cm and 2.0cm,
    scale=0.92,
    transform shape
  ]
    % --- Style definitions ---
    \tikzset{
     basicbox/.style = {draw, rounded corners, minimum height=1cm, align=center, font=\small, drop shadow={opacity=0.25, shadow xshift=2pt, shadow yshift=-2pt}},
     input/.style = {basicbox, fill=green!10, text width=2.2cm},
     process/.style = {basicbox, rectangle, fill=blue!10, text width=3.2cm, minimum height=1.4cm},
     secret/.style = {basicbox, fill=red!10, dashed, text width=4.5cm, font=\bfseries},
     artifact/.style = {basicbox, fill=yellow!20, double, text width=3.5cm},
     pqcbox/.style = {basicbox, dashed, fill=purple!10, text width=2cm},
     arrow/.style = {thick, ->, >=Stealth, rounded corners},
     lbl/.style = {fill=white, inner sep=2pt, font=\scriptsize, align=center, text=black!80, rounded corners=2pt}
    }

    % --- 1. Root Entropy ---
    \node (rev) [secret] {Root Entropy Value\\(REV - 256 bit)\\\normalfont\scriptsize\textit{Ephemeral / Never Stored}};

    % --- 2. Left Pipeline: Identity Derivation ---
    \node (hkdf) [process, below left=1.8cm and 2.0cm of rev] {\textbf{HKDF-SHA256}\\(Domain Separation)};

    % Identity Keys
    \node (ed25519) [basicbox, below=1.8cm of hkdf, xshift=-3.2cm, text width=2cm] {Ed25519\\Keypair};
    \node (secp) [basicbox, right=0.25cm of ed25519, text width=2cm] {Secp256k1\\Keypair};
    \node (x25519) [basicbox, right=0.25cm of secp, text width=2cm] {X25519\\(xID)};
    \node (dots) [right=0.1cm of x25519] {\Large \textbf{...}};
    \node (pqc) [pqcbox, right=0.1cm of dots, text width=2cm] {PQC\\Keypair};

    % --- 3. Right Pipeline: Sealing & Recovery ---
    \node (aes) [process, below right=1.8cm and 3.5cm of rev] {\textbf{AES-GCM-SIV}\\(Encryption)};
    \node (argon) [process, above=1.5cm of aes, fill=orange!20] {\textbf{Argon2id}\\(Key Derivation)};
    \node (pass_user) [input, above=1.0cm of argon, xshift=-2.8cm] {User\\Passphrase};
    \node (pass_admin) [input, above=1.0cm of argon, xshift=2.8cm] {Administrator\\Passphrase};
    \node (mnemonic) [artifact, below=1.2cm of aes] {\textbf{Sealed Artifact $\sigma$}\\(Encoded Ciphertext)};

    % --- Connections ---
    \draw [arrow] (rev) -| node[lbl, pos=0.3] {Raw Entropy\\(Context Input)} (hkdf);
    \draw [arrow] (hkdf) -- node[lbl, pos=0.5] {Derived\\Seeds} (ed25519.north);
    \draw [arrow] (hkdf) -- (secp.north);
    \draw [arrow] (hkdf) -- (x25519.north);
    \draw [arrow] (hkdf) -- (pqc.north);

    \draw [arrow] (pass_user) -- node[lbl, left, pos=0.4] {Passphrase\\Composition} (argon.north west);
    \draw [arrow] (pass_admin) -- (argon.north east);
    \draw [arrow] (argon) -- node[lbl] {Derived Sealing Key\\($K$)} (aes);

    \coordinate (turn1) at ($(rev.south) + (0, -0.8)$);
    \coordinate (turn2) at ($(aes.west) + (-1.0, 0)$);
    \draw [arrow] (rev.south) -- (turn1)
    -- node[lbl, pos=0.5, align=center, yshift=0.3cm] {Plaintext Payload\\(REV)}
    (turn1 -| turn2)
    -- (turn2)
    -- (aes.west);

    \draw [arrow] (aes) -- node[lbl] {Ciphertext + Tag\\(Sealed)} (mnemonic);

    % --- Background Grouping ---
    \begin{scope}[on background layer]
     \node [fit=(hkdf) (ed25519) (pqc), fill=blue!5, rounded corners, draw=blue!20, dashed, inner sep=0.4cm,
    label={[anchor=south west, blue!50, align=left]north west:\textbf{Pipeline 1:}\\Deterministic Identity\\(ADI Construction)}] {};
     \node [fit=(pass_user) (pass_admin) (argon) (mnemonic) (aes), fill=red!5, rounded corners, draw=red!20, dashed, inner sep=0.4cm,
    label={[red!50, align=center]above:\textbf{Pipeline 2:}\\Sealing \& Lifecycle}] {};
    \end{scope}
  \end{tikzpicture}%
  }
  \caption{\textbf{ACE-GF System Architecture.} \\ Arrows indicate data flow semantics. The design separates deterministic identity generation (left) from the secure sealing lifecycle (right). The architecture enforces that identity depends solely on REV, whereas authorization depends solely on passphrase binding.}
  \label{fig:architecture}
\end{figure}

\subsection{Identity Root Generation and Sealing}

\paragraph{Identity Root Generation.}
During initialization, the system samples a uniformly random 256-bit value
\[
\textsf{REV} \leftarrow \{0,1\}^{256},
\]
which serves as the identity root. The REV is never persisted or exported and exists only ephemerally in memory during execution.

\paragraph{Sealing Key Derivation.}
An authorization credential $\textsf{Cred}$ (e.g., a user- or administrator-supplied passphrase) is transformed into a high-entropy symmetric sealing key
\[
K_{\text{seal}} \leftarrow \textsf{Argon2id}(\textsf{Cred}, s),
\]
using a memory-hard key derivation function, where $s$ is a 128-bit random salt sampled independently for each sealed artifact and stored with it in the clear. This step raises the computational cost of offline guessing attacks against the sealed artifact.

\paragraph{Sealing with a Fixed Nonce.}
The identity root $\textsf{REV}$ is encrypted under $K_{\text{seal}}$ using AES-GCM-SIV, producing a sealed artifact $\textsf{SA}$:
\[
\textsf{SA} \leftarrow \bigl(s,\ \textsf{Enc}_{K_{\text{seal}}}(N_{\text{fixed}}, \textsf{REV})\bigr).
\]
A fixed, public nonce $N_{\text{fixed}}$ is used for all sealing operations. The nonce-misuse resistance of AES-GCM-SIV ensures that confidentiality of the plaintext is preserved even under nonce reuse, provided the sealing key remains secret. This design choice removes nonce state; the only randomness in sealing is the per-artifact salt $s$.

\paragraph{Storage Model.}
Only the sealed artifact $\textsf{SA} = (s, c)$ is stored persistently, for example as a mnemonic encoding or as an opaque ciphertext blob. The REV is securely erased from memory immediately after sealing. At no point does unencrypted identity root material exist at rest.

\paragraph{Wallet Encoding Profile.}\label{sec:wallet-profile}
The reference implementation uses an encoding profile that matches the 24-word recovery phrases wallet users already keep: the sealed artifact is written as 24 words from the BIP-39 English wordlist. Only the wordlist and checksum are reused; the phrase is not a BIP-39 seed and does not work in BIP-39 wallets. Twenty-four words hold 32 bytes, so the profile samples 16 bytes of identity material (128-bit source entropy) and stores its 16-byte AES-GCM-SIV ciphertext together with the 16-byte tag; HKDF expands the material into a 32-byte identity root after unsealing. No room is left for a per-artifact salt, a version, or KDF parameters. Argon2id therefore uses a fixed public salt, and one guess tests the same credential against every artifact an adversary holds (Section~\ref{sec:sa-confidentiality}). A deployment with a stable public identifier may pass it as the Argon2id salt. With 128-bit source entropy the root reaches at most NIST post-quantum Category~1 (comparable to AES-128 key search), against Category~5 for a 256-bit REV, and a search over $U$ users' public keys lowers this by a factor of about $U$ classically and about $\sqrt{U}$ for Grover search. The general construction (256-bit REV, per-artifact salt) is unchanged; the profile gives up these properties to keep the 24-word format.

\subsection{Context-Isolated Key Derivation}

All cryptographic keys in ACE-GF are deterministically derived from the REV using HKDF with explicit context encoding. This mechanism enforces cryptographic isolation across algorithms, protocols, and application domains.

\paragraph{Context Structure.}
Each derivation is parameterized by a structured context tuple
\[
\textsf{Ctx} = (\textsf{AlgID}, \textsf{Domain}, \textsf{Index}),
\]
where:
\begin{itemize}
  \item $\textsf{AlgID}$ identifies the target cryptographic algorithm or curve (e.g., Ed25519, Secp256k1, Dilithium),
  \item $\textsf{Domain}$ specifies the application domain (e.g., signing, encryption, authentication),
  \item $\textsf{Index}$ provides an application-specific derivation index.
\end{itemize}

\paragraph{Derivation Procedure.}
Key derivation proceeds as follows:
\[
\textsf{PRK} \leftarrow \textsf{HKDF-Extract}(\textsf{salt}, \textsf{REV}),
\]
\[
\textsf{Key} \leftarrow \textsf{HKDF-Expand}(\textsf{PRK}, \textsf{Ctx}, L).
\]
Here $\textsf{salt}$ is the absent salt of RFC~5869, a string of 32 zero bytes: $\textsf{REV}$ is already uniform, so extraction needs no secret salt. It is unrelated to the per-artifact sealing salt $s$.
The context tuple is supplied as the \texttt{info} parameter to HKDF-Expand through an injective encoding $\Encode$ of the tuple, for example with each variable-length field prefixed by its length, so that distinct context tuples yield distinct \texttt{info} strings and keys derived under distinct contexts are computationally independent. Plain concatenation of variable-length fields is not injective and does not give this guarantee.

\paragraph{Isolation Guarantee.}
Under the pseudorandomness assumption of HKDF, any two keys derived with different context tuples are indistinguishable from random and independent of one another. Consequently, compromise of a key in one algorithm or application domain does not enable recovery of keys in other domains, nor does it reveal information about the underlying REV.

\paragraph{Post-Quantum Cryptographic Readiness.}
Post-quantum algorithms are added through the same context mechanism. When a post-quantum algorithm (e.g., Dilithium) is introduced, a new $\textsf{AlgID}$ value is used. This yields an independent derivation domain without modifying the identity root, sealed artifact, or any previously derived classical keys, so migration does not disrupt existing keys.

\subsection{Identity Lifecycle Management}

The separation between identity and authorization enables lifecycle operations that do not require rekeying or persistent state.

\paragraph{Stateless Credential Rotation.}
To update authorization credentials, the system first reconstructs the identity root by executing
\[
\textsf{REV} \leftarrow \textsf{Unseal}(\textsf{SA}, \textsf{Cred}_{\text{old}}),
\]
and then re-seals the same REV under a new credential:
\[
\textsf{SA}' \leftarrow \textsf{Seal}(\textsf{REV}, \textsf{Cred}_{\text{new}}).
\]
All derived keys remain unchanged, as they depend solely on the REV. No server-side state or key regeneration is required.

\paragraph{Authorization-Bound Revocation.}
Authorization revocation is achieved by rendering the relevant credential component unavailable (e.g., deletion of an administrator-controlled secret). Since reconstruction of the REV requires possession of the complete credential material, revocation of future access takes effect without modifying identity-derived keys or maintaining revocation state, under the conditions stated below.

\paragraph{Scope of Rotation and Revocation.}
Rotation and revocation govern \emph{future} access through the credential path: they stop a party that has not already obtained the REV, or a sealed artifact together with its valid credential, from reconstructing the REV. They do not provide post-compromise security. The REV is invariant by design, because invariance is what preserves identity continuity. Any party that has previously held the REV, or a superseded sealed artifact together with its credential, can still derive every identity key. Rotation therefore assumes that superseded sealed artifacts or the old credential are destroyed, and revocation assumes that no copy of the removed component, or of a sealing key computed from it, remains. Recovery from an actual exposure of the REV requires a new REV, that is, a new identity and migration of its assets; changing the derivation context helps only when a single derived key, not the REV, was exposed.

Algorithms~\ref{alg:core} and~\ref{alg:lifecycle} consolidate the ACE-GF construction as formal pseudocode.

\begin{algorithm}[t]
\caption{ACE-GF Core Operations}\label{alg:core}
\begin{algorithmic}[1]
\Procedure{Setup}{$1^\lambda$}
  \State \textbf{select} $\Argon$ parameters $(m, t, p)$ for security level $\lambda$
  \State \textbf{fix} public nonce $N_{\text{fixed}}$
  \State \Return $\mathsf{params} \gets (\lambda,\, m,\, t,\, p,\, N_{\text{fixed}})$
\EndProcedure
\Statex
\Procedure{Seal}{$\mathsf{params},\; \mathsf{Cred}$}
  \State $\REV \stackrel{\$}{\leftarrow} \{0,1\}^{256}$ \Comment{sample identity root}
  \State $s \stackrel{\$}{\leftarrow} \{0,1\}^{128}$ \Comment{per-artifact salt}
  \State $K_{\text{seal}} \gets \Argon(\mathsf{Cred};\, s;\, m, t, p)$
  \State $c \gets \AESGCMSIV.\mathrm{Enc}(K_{\text{seal}},\; N_{\text{fixed}},\; \REV)$
  \State \textbf{zeroize} $K_{\text{seal}},\; \REV$
  \State \Return $\mathsf{SA} \gets (s,\, c)$
\EndProcedure
\Statex
\Procedure{Unseal}{$\mathsf{params},\; \mathsf{SA},\; \mathsf{Cred}$}
  \State \textbf{parse} $\mathsf{SA}$ as $(s,\, c)$
  \State $K_{\text{seal}} \gets \Argon(\mathsf{Cred};\, s;\, m, t, p)$
  \State $\REV \gets \AESGCMSIV.\mathrm{Dec}(K_{\text{seal}},\; N_{\text{fixed}},\; c)$
  \State \textbf{zeroize} $K_{\text{seal}}$
  \If{decryption failed}
    \State \Return $\perp$
  \EndIf
  \State \Return $\REV$ \Comment{ephemeral; caller must zeroize}
\EndProcedure
\end{algorithmic}
\end{algorithm}

\begin{algorithm}[t]
\caption{ACE-GF Derivation and Lifecycle}\label{alg:lifecycle}
\begin{algorithmic}[1]
\Procedure{Derive}{$\REV,\; \mathsf{Ctx} = (\mathsf{AlgID},\, \mathsf{Domain},\, \mathsf{Index})$}
  \State $\mathsf{PRK} \gets \HKDF\text{-}\mathrm{Extract}(\mathsf{salt},\; \REV)$ \Comment{$\mathsf{salt}=0^{256}$, RFC~5869 default}
  \State $\mathsf{info} \gets \Encode(\mathsf{AlgID},\, \mathsf{Domain},\, \mathsf{Index})$ \Comment{injective, length-prefixed}
  \State $\mathsf{Key} \gets \HKDF\text{-}\mathrm{Expand}(\mathsf{PRK},\; \mathsf{info},\; L)$
  \State \textbf{zeroize} $\mathsf{PRK}$
  \State \Return $\mathsf{Key}$
\EndProcedure
\Statex
\Procedure{RotateCredential}{$\mathsf{SA},\; \mathsf{Cred}_{\text{old}},\; \mathsf{Cred}_{\text{new}}$}
  \State $\REV \gets \Call{Unseal}{\mathsf{params},\; \mathsf{SA},\; \mathsf{Cred}_{\text{old}}}$
  \If{$\REV = \perp$}
    \State \Return $\perp$
  \EndIf
  \State $s' \stackrel{\$}{\leftarrow} \{0,1\}^{128}$
  \State $K'_{\text{seal}} \gets \Argon(\mathsf{Cred}_{\text{new}};\, s';\, m, t, p)$
  \State $c' \gets \AESGCMSIV.\mathrm{Enc}(K'_{\text{seal}},\; N_{\text{fixed}},\; \REV)$
  \State \textbf{zeroize} $K'_{\text{seal}},\; \REV$
  \State \Return $\mathsf{SA}' \gets (s',\, c')$
\EndProcedure
\end{algorithmic}
\end{algorithm}

\subsection{Parametric Agility for Constrained Environments}

A common critique of memory-hard functions is their computational demand on resource-constrained IoT hardware. ACE-GF addresses this through parametric agility: the Argon2id parameters $(m, t, p)$ can be tuned based on the device's available SRAM and power profile.

In scenarios where hardware-backed trust anchors (e.g., ARM TrustZone or Secure Enclaves) are present, the memory-hardness requirements can be relaxed only if the sealed artifact $\textsf{SA}$ never leaves the hardware boundary. Lower Argon2id parameters lower the cost of each offline guess by the same factor, so an artifact that is extracted from the device, or backed up outside it, is then protected only by the credential's entropy at the reduced cost (Section~\ref{sec:sa-confidentiality}). The sealing cost can thus range from a high $m_{cost}$ for passphrase protection to minimal overhead for hardware-entangled identities; identity derivation is the same in every case.

%%%%%%%%%%%%%%%%%%%%%%%%%%%%%%%%%%%%%%%%%%%%%%%%%%%%%%%%%%%%
%        5. Security Analysis and Arguments                %
%%%%%%%%%%%%%%%%%%%%%%%%%%%%%%%%%%%%%%%%%%%%%%%%%%%%%%%%%%%%

\section{Security Analysis and Arguments}

This section provides security arguments for ACE-GF under standard cryptographic assumptions. We introduce no new security notions or primitives, and focus on the confidentiality of the sealed artifact and the isolation of derived keys across distinct cryptographic contexts.

\subsection{Security Model and Assumptions}

Our security arguments rely on the following widely accepted assumptions:
\begin{itemize}
  \item \textbf{HKDF-SHA256.}
  HKDF-SHA256 is modeled as a secure pseudorandom function (PRF) when instantiated with a cryptographic hash function and analyzed in the random oracle model.

  \item \textbf{AES-GCM-SIV.}
  AES-GCM-SIV is assumed to be a secure misuse-resistant authenticated encryption (MRAE) scheme, preserving confidentiality and integrity even under nonce reuse.
\end{itemize}

We further assume that authorization credentials supplied to the system possess sufficient entropy for their intended threat model, and that standard cryptographic implementations are used without introducing side-channel vulnerabilities beyond those discussed in Section~5.4.

\subsection{Confidentiality of the Sealed Artifact}\label{sec:sa-confidentiality}

The sealed artifact $\textsf{SA}$ encodes the identity root $\textsf{REV}$ in encrypted form and is the only persistent object stored by the system. We analyze its confidentiality under the misuse-resistant authenticated encryption (MRAE) notion.

\begin{claim}[Confidentiality of Sealed Artifacts]
The sealed artifact produced by ACE-GF is computationally indistinguishable from random to any probabilistic polynomial-time adversary without knowledge of the correct authorization credential, up to the offline-guessing term $q/2^{\mu}$ for $q$ Argon2id evaluations against a credential of min-entropy $\mu$ (bound below).
\end{claim}

\paragraph{Argument Sketch.}
The identity root $\textsf{REV}$ is encrypted using AES-GCM-SIV under a sealing key $K_{\text{seal}}$ derived from the authorization credential via Argon2id. A fixed, public nonce is used for all sealing operations. Due to the nonce-misuse resistance of AES-GCM-SIV, confidentiality of the ciphertext is preserved even under nonce reuse, provided that the sealing key remains secret.

Recovering $\textsf{REV}$ from $\textsf{SA}$ therefore requires either breaking the MRAE security of AES-GCM-SIV or guessing the correct sealing key. Since $K_{\text{seal}}$ is derived from the credential using a memory-hard function, offline guessing attacks are computationally expensive and bounded by the entropy of the credential. Under the stated assumptions, the sealed artifact does not leak information about the identity root.

\paragraph{Offline Guessing Bound.}
The claim holds up to an unavoidable offline-guessing term. Model Argon2id as a random oracle and let $\mu$ be the min-entropy of the credential. An adversary making $q$ Argon2id evaluations distinguishes a given artifact from random, or recovers its REV, with advantage at most $q/2^{\mu} + \mathsf{Adv}^{\mathrm{mrae}}_{\textsf{AES-GCM-SIV}}$. With per-artifact salts, the probability of compromising at least one of $U$ artifacts stays at about $q/2^{\mu}$ plus the multi-user MRAE term, because each evaluation tests a single artifact. Under the wallet encoding profile (Section~\ref{sec:wallet-profile}) the salt is a fixed public constant, so each evaluation tests all $U$ artifacts at once: the probability of compromising at least one of them is at most $qU/2^{\mu} + U\cdot\mathsf{Adv}^{\mathrm{mrae}}_{\textsf{AES-GCM-SIV}}$, and the evaluations can be computed before any artifact is obtained. Under that profile the sealed artifact is a second factor that its holder keeps confidential, not a public object.

\subsection{Context Isolation of Derived Keys}

All cryptographic keys in ACE-GF are derived from the identity root using HKDF with explicit context encoding. We analyze the isolation guarantees provided by this derivation mechanism.

\begin{claim}[Context Isolation]
Any two keys derived under distinct context descriptors are computationally independent, even if an adversary obtains full knowledge of one of the derived keys.
\end{claim}

\paragraph{Argument Sketch.}
Derived keys are computed using HKDF-Expand with the context descriptor supplied as the \texttt{info} parameter. Because $\Encode$ is injective, distinct context descriptors give distinct \texttt{info} values. Under the PRF assumption for HKDF, outputs corresponding to distinct \texttt{info} values are pseudorandom and independent. Consequently, knowledge of a derived key associated with one algorithm, domain, or application context does not enable computation of keys derived under different contexts, nor does it allow computing the underlying identity root: an adversary that recovered $\textsf{REV}$ from one derived key could compute the key for any other context and so distinguish HKDF from a random function.

Classical and post-quantum keys derived from the same identity root can therefore be used in parallel without affecting one another.

\subsection{Security Boundaries and Limitations}

The security guarantees provided by ACE-GF are subject to the following limitations:

\paragraph{Credential Entropy.}
The confidentiality of the sealed artifact depends on the entropy of the authorization credential. While Argon2id significantly increases the cost of offline guessing attacks, the construction cannot prevent compromise if low-entropy credentials are used against a sufficiently powerful adversary.

\paragraph{Side-Channel Exposure.}
The security model does not account for advanced side-channel attacks against the identity root while it exists ephemerally in memory, such as cold-boot attacks or fine-grained timing analysis. Deployments with elevated threat models should execute unsealing and key derivation within trusted execution environments (TEEs), as discussed in Section~6.5.

Both limitations apply to software-based identity systems in general and are independent of the cryptographic structure of ACE-GF.

%%%%%%%%%%%%%%%%%%%%%%%%%%%%%%%%%%%%%%%%%%%%%%%%%%%%%%%%%%%%
%    6. Application, Implementation, and Deployment        %
%%%%%%%%%%%%%%%%%%%%%%%%%%%%%%%%%%%%%%%%%%%%%%%%%%%%%%%%%%%%

\section{Application, Implementation, and Deployment}

This section discusses practical implementation considerations, deployment environments, and migration capabilities enabled by ACE-GF. We focus on portability and interoperability, not on performance optimization for a specific platform.

\subsection{Reference Implementation and Determinism}

We provide a deterministic reference implementation of ACE-GF written in Rust. The implementation serves as a proof-of-concept and validation of the construction described in Sections~3 and~4, instantiated with the wallet encoding profile of Section~\ref{sec:wallet-profile}. All security claims are derived from the formal specification and underlying cryptographic assumptions; the reference implementation itself is not part of the security model.

The core library is implemented in \texttt{no\_std}-compatible Rust and provides the following functionality:
\begin{itemize}
  \item sampling of the identity material (16 bytes, expanded by HKDF to a 32-byte identity root, under the wallet encoding profile of Section~\ref{sec:wallet-profile}),
  \item AES-GCM-SIV sealing and unsealing of the identity material,
  \item HKDF-SHA256 expansion with explicit and fixed context encoding,
  \item deterministic key derivation for Ed25519, Secp256k1, and X25519,
  \item credential binding and authorization-bound revocation mechanisms.
\end{itemize}

All sensitive values, including the REV, intermediate seeds, sealing keys, and derived private keys, exist only ephemerally in memory and are explicitly zeroized upon leaving scope. Stack-local buffers and the \texttt{zeroize} trait are used to minimize memory remanence.

\subsection{Portability and Cross-Platform Consistency}

The reference implementation has been compiled for the following targets:
\begin{itemize}
  \item native platforms via C-compatible FFI bindings,
  \item WebAssembly targets,
  \item iOS and Android mobile environments.
\end{itemize}

For a fixed sealed artifact and authorization credential, all supported platforms produce byte-for-byte identical outputs. This determinism holds across architectures and endianness, as nonce formats, context labels, and encoding conventions are fully specified and fixed. No platform persists the identity root or any derived secret material at rest.

\subsection{Verification Artifacts and Proof-of-Concept}

The public reference repository provides auxiliary artifacts intended to support verification, experimentation, and independent reproduction of the scheme's behavior. These artifacts are illustrative and are not intended as production-ready software.

\paragraph{Command-Line Interface.}
A minimal command-line interface demonstrates core operations, including identity generation, deterministic reconstruction, stateless credential rotation, and inspection of derived identity addresses across cryptographic domains.

\paragraph{Test Suite.}
Unit tests check the correctness of the cryptographic core and system invariants, including:
\begin{itemize}
  \item cryptographic independence across context-isolated derivations,
  \item correct handling of composed authorization credentials for dual-control and revocation scenarios,
  \item Diffie--Hellman key agreement and end-to-end encryption key resolution,
  \item round-trip migration of pre-existing private keys into a deterministically anchored identity root.
\end{itemize}

\paragraph{Determinism Vectors.}
A set of standardized test vectors, encoded in JSON format, validates strict deterministic behavior across platforms. These vectors include profiles using single-factor authorization credentials as well as inputs containing complex UTF-8 characters, such as non-Latin scripts and emoji, to check that credentials are processed identically across platforms.

The reference implementation and associated artifacts are available at:
\begin{center}
\url{https://github.com/ya-xyz/acegf-playground}
\end{center}

\subsection{Performance Rationale and Optimization}

The computational cost of ACE-GF is primarily localized to the initial \textit{Unseal} operation. Once the REV is reconstructed ephemerally, subsequent key derivations via HKDF-SHA256 are computationally negligible~\cite{RFC5869, rogaway2006}.

On IoT devices, the REV can be reconstructed once during the device boot sequence or user-session initialization and held in zeroizable volatile memory. This limits the Argon2id overhead to a single ``cold start'' event, at the cost of keeping the REV in memory for the whole session rather than for a single operation; the exposure to memory attacks discussed in Section~5.4 grows accordingly. On sub-watt devices, the credential-intensive key derivation can be delegated to a trusted pairing device (e.g., a mobile gateway), allowing the end-node to perform only the lightweight AES-GCM-SIV decryption to achieve identity continuity. The pairing device then sees the credential and computes $K_{\text{seal}}$, which decrypts the sealed artifact without the credential; it is therefore inside the trust boundary of the identity, and a copy of $K_{\text{seal}}$ retained by it remains valid until the artifact is re-sealed under a new credential or salt.

\subsection{Deployment and Migration Considerations}

ACE-GF is designed as a software-level identity construction that does not rely on centralized infrastructure or specialized hardware. The construction can be deployed in fully decentralized environments while preserving deterministic identity reconstruction and strict cryptographic isolation. When higher assurance is required, the unsealing and key derivation operations may be executed within trusted execution environments (TEEs), such as Intel SGX or ARM TrustZone, to mitigate the side-channel risks discussed in Section~5.4.

Explicit context isolation in the derivation process also supports interoperability and long-term deployment. Introduction of new cryptographic algorithms, including post-quantum primitives, requires only the assignment of fresh context identifiers. This allows classical and post-quantum keys to coexist under a single stable identity root without re-enrollment, rekeying of existing domains, or modification of stored sealed artifacts.

ACE-GF also supports migration from existing systems. Legacy private keys can be incorporated by encrypting them under a dedicated wrapping key derived from the identity root using a reserved context label. Legacy and newly derived keys can then be managed under one identity root, and legacy identities keep their original semantics. The wrapped key is an additional stored ciphertext, and it keeps whatever exposure history it had before migration.

\paragraph{Legacy Key Embedding (Optional).}
Although ACE-GF is defined as a seed-storage-free identity construction, practical deployments frequently contain long-lived legacy private keys, such as ECDSA keys controlling existing blockchain assets. For such deployments, we provide an optional helper mechanism that embeds such keys into an ACE-GF identity without modifying the construction or affecting the security model.

Given a legacy private key $\mathsf{sk}_{\mathrm{legacy}}$ and an authorization credential $\mathsf{Cred}$, the helper function
\[
\texttt{wei\_to\_crypto\_entity}
:\;
(\mathsf{sk}_{\mathrm{legacy}}, \mathsf{Cred})
\longrightarrow
\mathsf{SA}
\]
produces a sealed artifact whose plaintext is the legacy key material, so that the legacy address is reproduced from it, and uses that material as the identity root for the other context-isolated keys.

This helper serves solely as a practical migration layer, and it lies outside the security analysis of Section~5. The embedded root is not sampled by $\textsf{Seal}$, may have been exposed before migration, and is used directly as a key in the legacy context: exposure of the legacy key therefore exposes the identity root and every key derived from it, so the context-isolation claim does not cover such an identity. The helper also needs room for the full legacy key in the sealed artifact. It applies to the general construction or other encodings with at least the key length of plaintext, and not to the 24-word wallet encoding profile (Section~\ref{sec:wallet-profile}), whose 16 bytes of material cannot carry a 32-byte legacy key.

%%%%%%%%%%%%%%%%%%%%%%%%%%%%%%%%%%%%%%%%%%%%%%%%%%%%%%%%%%%%
%            7. Conclusion and Future Work                  %
%%%%%%%%%%%%%%%%%%%%%%%%%%%%%%%%%%%%%%%%%%%%%%%%%%%%%%%%%%%%

\section{Conclusion and Future Work}

We presented ACE-GF, a seed-storage-free identity construction for deterministic key management. It combines misuse-resistant authenticated encryption with context-isolated key derivation, so the identity root can be reconstructed deterministically while no master secret is stored at rest. Credentials can be rotated or revoked without changing the identity, although both operations govern future access only. Explicit context encoding isolates signing, encryption, authentication, and asset-control keys, and lets classical and post-quantum algorithms coexist under the same identity root.

The construction uses only well-established cryptographic primitives and inherits their security properties under standard assumptions. Because the identity root exists only ephemerally in memory and is never stored or exported, ACE-GF reduces the impact of key compromise and behaves consistently across platforms and trust boundaries.

A formal protocol specification of ACE-GF has been submitted as an IETF Internet-Draft~\cite{draft-acegf}, providing implementors with precise data structures, encoding conventions, and test vectors for interoperable deployment.

\paragraph{Future Work.}
Future work has two main directions. First, we plan to develop formal security proofs in the standard model, reducing reliance on random oracle assumptions and strengthening assurance for deployment in composable systems. Second, we intend to evaluate the construction under practical deployment conditions, including empirical side-channel analysis of the \textsf{Unseal} and \textsf{Derive} operations on real trusted execution environment (TEE) hardware, and to design corresponding mitigation strategies.

Additional directions include systematic integration of standardized post-quantum signature and key encapsulation mechanisms within the context-isolated derivation framework, as well as exploration of protocol-level applications for machine-to-machine identity, autonomous agent coordination, and long-lived digital services.

Future work will also involve benchmarking ACE-GF on specific ARM Cortex-M architectures and refining the adaptive parameter selection algorithms to automatically balance brute-force resistance against power consumption in decentralized autonomous sensor networks.

%%%%%%%%%%%%%%%%%%%%%%%%%%%%%%%%%%%%%%%%%%%%%%%%%%%%%%%%%%%%
%                     References                           %
%%%%%%%%%%%%%%%%%%%%%%%%%%%%%%%%%%%%%%%%%%%%%%%%%%%%%%%%%%%%

%%%%%%%%%%%%%%%%%%%%%%%%%%%%%%%%%%%%%%%%%%%%%%%%%%%%%%%%%%%%
%                     Appendix A                           %
%%%%%%%%%%%%%%%%%%%%%%%%%%%%%%%%%%%%%%%%%%%%%%%%%%%%%%%%%%%%

\appendix
\section{Non-Normative Applications}

This appendix describes illustrative application scenarios enabled by the ACE-GF construction. The material in this section is \emph{non-normative} and is not part of the formal system model, construction, or security analysis.

\subsection{Multi-Curve Identity and Asset Control}

A single Root Entropy Value ($\mathsf{REV}$) deterministically anchors cryptographic material across heterogeneous ecosystems, including signature, key-agreement, and encryption domains (e.g., Ed25519, Secp256k1, X25519). Through explicit context encoding, each derived key is isolated from all others while remaining reconstructible from the same identity root.

One identity can therefore control assets and authenticate across several cryptographic environments and platforms without separate master secrets.

\subsection{Authorization-Controlled Lifecycles}

Access to the identity root is governed by authorization credentials that are separate from the identity, so credentials can change over time without persistent secret storage.

Credential updates and revocation can be expressed by modifying or removing designated authorization components, without requiring changes to the underlying identity root or re-derivation of existing cryptographic domains. This model supports organizational key management workflows in which identity continuity must be preserved while authorization policies change.

\end{document}